\documentclass[conference]{IEEEtran}
\IEEEoverridecommandlockouts
\usepackage{cite}
\usepackage{amsmath,amssymb,amsfonts}
\usepackage{algorithmic}
\usepackage{graphicx}
\usepackage{textcomp}
\usepackage{xcolor}
\usepackage{float}
\usepackage{booktabs}
\def\BibTeX{{\rm B\kern-.05em{\sc i\kern-.025em b}\kern-.08em
    T\kern-.1667em\lower.7ex\hbox{E}\kern-.125emX}}
\begin{document}

\title{CardioPatternFormer: Pattern-Guided Attention for Interpretable ECG Classification with Transformer Architecture\\

}

\author{\IEEEauthorblockN{1\textsuperscript{st} Berat Kutay Uğraş}
\IEEEauthorblockA{\textit{Department of Electrical and Electronics Engineering} \\
\textit{Eskisehir Technical University}\\
Eskisehir, Turkey \\
ORCID: 0009-0006-3760-4870}
\and

\IEEEauthorblockN{2\textsuperscript{nd} İbrahim Talha Saygı}
\IEEEauthorblockA{\textit{Department of Electrical and Electronics Engineering} \\
\textit{Eskisehir Technical University}\\
Eskisehir, Turkey \\
email address or ORCID}
\and 
\IEEEauthorblockN{3\textsuperscript{rd} Ömer Nezih Gerek}
\IEEEauthorblockA{\textit{Department of Electrical and Electronics Engineering} \\
\textit{Eskisehir Technical University}\\
Eskisehir, Turkey \\
ORCID: 0000-0001-8183-1356}

}

\maketitle

\begin{abstract}
Electrocardiogram (ECG) interpretation is fundamental to cardiac diagnosis, but deep learning models often lack transparency, hindering clinical trust. We introduce CardioPatternFormer, a novel transformer-based architecture that reframes ECG interpretation through the lens of pattern recognition, treating cardiac patterns as a vocabulary learned from data. CardioPatternFormer integrates several innovations: (1) a Cardiac Pattern Tokenizer that decomposes ECG signals into learned, multi-scale patterns; (2) Physiologically Guided Attention mechanisms incorporating adaptable, domain-specific constraints based on cardiac electrophysiology; (3) Multi-Resolution Temporal Encoding to capture diverse temporal dynamics; and (4) specialized classification heads providing class-specific attention visualizations for detailed diagnostic explanations. Evaluated on the Chapman-Shaoxing dataset across major diagnostic categories, CardioPatternFormer demonstrates strong classification performance, particularly for rhythm disorders, with results aligning with clinical experience regarding diagnostic difficulty gradients. More significantly, CardioPatternFormer enhances interpretability by visualizing physiologically relevant ECG regions influencing each diagnosis, bridging automated analysis and clinical reasoning. This pattern-centric approach advances ECG classification and establishes a foundation for more transparent and clinically integrated cardiac signal analysis.
\end{abstract}

\renewcommand{\IEEEkeywordsname}{Keywords}

\section{INTRODUCTION}
The electrocardiogram (ECG) stands as a cornerstone of cardiovascular diagnostics, offering a non-invasive window into the heart's electrical activity. Annually, hundreds of millions of ECGs are performed worldwide, underscoring its fundamental role in identifying a wide spectrum of cardiac conditions \cite{20}. Despite its ubiquity and the wealth of information it provides, accurate ECG interpretation demands considerable expertise, typically cultivated over years of rigorous training and clinical practice. Even amongst seasoned cardiologists, inter-reader variability remains a notable challenge, with studies reporting significant discrepancies in the identification of specific cardiac abnormalities. This variability highlights not only the inherent complexities of ECG analysis but also the persistent need for advanced computational tools that can deliver consistent, precise, and interpretable analyses.

Deep learning (DL) methodologies have demonstrated remarkable potential in automating ECG interpretation, with several studies showcasing their ability to achieve cardiologist-level performance for specific diagnostic tasks. However, the clinical adoption of these powerful algorithms has been hampered by two primary limitations. Firstly, many DL models operate as "black boxes," offering limited insight into their decision-making processes, which can erode clinical trust and hinder their integration into routine practice. Secondly, these models often distill the rich physiological data embedded within ECG signals into simplified binary or categorical outputs, thereby losing the nuanced patterns and temporal relationships that clinicians utilize for comprehensive diagnostic reasoning.

We propose that ECG signals can be conceptualized as a complex "language" of cardiac electrophysiology. This language comprises discrete, recognizable elements (the "vocabulary") which combine in specific temporal sequences (the "grammar") to form a complete diagnostic picture. Just as human languages convey meaning through words and syntax, cardiac pathologies manifest through distinct patterns and their temporal interplay within the ECG. This perspective suggests that the challenge of ECG interpretation can be effectively addressed through advanced pattern recognition and contextual understanding, areas where transformer architectures, initially popularized in natural language processing \cite{2}, have shown exceptional promise.

In this paper, we introduce CardioPatternFormer, a novel framework for ECG interpretation that leverages a physiologically-informed pattern tokenization approach coupled with specialized attention mechanisms. Our primary contributions are:

\begin{itemize}
     \item A CardiacPatternTokenizer that transforms continuous ECG signals into a vocabulary of learned cardiac patterns, capturing salient electrophysiological features across multiple time scales.
    \item Physiologically Guided Attention, a novel mechanism that refines standard attention scores by incorporating adaptable biases derived from established principles of cardiac electrophysiology, such as signal locality and rhythmic characteristics. The influence of these biases is governed by learnable weights, thereby guiding the model to focus more effectively on diagnostically relevant signal features.
     \item A Multi-Resolution Temporal Encoding scheme designed to capture both localized morphological abnormalities and broader rhythm patterns, which are critical for comprehensive ECG interpretation.
     \item Class-specific attention visualization derived from specialized classification heads, providing detailed, interpretable explanations for each diagnostic prediction, thereby addressing the critical need for explainability in clinical applications.
\end{itemize}

CardioPatternFormer demonstrates strong classification performance on the publicly available Chapman-Shaoxing ECG dataset \cite{3} across six major diagnostic categories, with notable strength in identifying complex rhythm disorders and conduction abnormalities. More significantly, it offers enhanced transparency in its decision-making process through visualizations that highlight the specific ECG regions influencing each diagnosis, thereby creating a valuable bridge between automated analysis and clinical reasoning. Furthermore, we evaluate the model’s adaptability to varying lead availability through a comprehensive lead ablation study.

While current approaches to automated ECG interpretation primarily frame the task as classification, the pattern vocabulary and attention mechanisms developed in this work also establish a foundation for more sophisticated diagnostic understanding. The foundational pattern recognition capabilities established here may support future work exploring automated report generation or other advanced diagnostic communication methods.

The remainder of this paper is organized as follows: Section II discusses related work in deep learning for ECG interpretation and attention-based explainability. Section III details the CardioPatternFormer architecture and its components. Section IV presents our experimental setup and results. Section V discusses the implications of our approach, its clinical potential, limitations, and future directions, followed by the Conclusion in Section VI.

\section{RELATED WORK}

\subsection{Deep Learning for ECG Interpretation}

The application of deep learning to ECG interpretation has witnessed substantial growth and evolution over the past decade. Initial forays predominantly relied on Convolutional Neural Networks (CNNs) to automatically extract morphological features from ECG waveforms. Deep CNN architectures demonstrated notable efficacy in arrhythmia detection, particularly from single-lead ECGs, with some studies reporting performance comparable to that of experienced cardiologists for specific rhythm disorders \cite{1}. This paradigm was subsequently extended to 12-lead ECG analysis, often employing advanced CNN structures like Residual Networks (ResNets) \cite{4}. Furthermore, researchers have successfully utilized CNNs to identify subtle ECG indicators of conditions such as left ventricular dysfunction, uncovering previously unrecognized signal patterns with significant diagnostic value \cite{5}.

To address the temporal dependencies inherent in ECG data, Recurrent Neural Networks (RNNs), especially Long Short-Term Memory (LSTM) networks, have been extensively applied. These models are well-suited for capturing sequential information and have been used for tasks ranging from arrhythmia classification to the prediction of adverse cardiac events \cite{6, 7}. Hybrid models, which integrate the feature extraction strengths of CNNs with the sequential modeling capabilities of RNNs (often termed CNN-RNN architectures), have also emerged as a popular approach, demonstrating improved performance in complex tasks like multi-label arrhythmia detection from extended ECG recordings \cite{8}. More recently, self-supervised pre-training on large, unlabeled ECG datasets has shown promise in learning robust and generalizable feature representations, which can then be fine-tuned for specific diagnostic tasks with enhanced data efficiency \cite{9}. Despite these significant advances, a persistent challenge across many of these models is their limited interpretability, often treating ECG analysis as a complex pattern-matching problem without explicitly modeling underlying physiological mechanisms.

\subsection{Transformer Architectures for Physiological Signals}

Transformer architectures, originally introduced for natural language processing tasks \cite{2}, have recently garnered significant attention for their potential in analyzing physiological time series, including ECGs. Early applications of transformers to ECG interpretation demonstrated their ability to effectively model long-range dependencies within the signal, leading to improved performance in arrhythmia detection compared to conventional CNN and RNN approaches \cite{10}. Subsequent research has focused on adapting transformer architectures more specifically for ECG data, for instance, by incorporating relative positional encodings to better capture the temporal relationships crucial for cardiac signal analysis. General-purpose transformer-based architectures for time series, such as TimesNet, have also shown strong performance on ECG classification benchmarks \cite{11}. However, many existing methods apply transformer architectures with minimal domain-specific modifications, often treating ECG signals as generic time-series data rather than as reflections of complex cardiac electrophysiology. A key limitation in these approaches often lies in the tokenization strategy, where individual time points or fixed-size windows are used as tokens, which may not align optimally with physiologically meaningful segments or events within the ECG.

\subsection{Explainability in Cardiac Models}

The integration of deep learning models into clinical decision-making processes necessitates a high degree of transparency and interpretability. This has spurred significant research into eXplainable Artificial Intelligence (XAI) for cardiac applications. In ECG interpretation, several techniques have been explored to elucidate model predictions. Gradient-based visualization methods, such as Grad-CAM \cite{12} and its variants, have been widely applied to generate saliency maps that highlight the regions of the ECG signal most influential in a model's decision \cite{13}. Such visualizations can aid clinicians in understanding the basis of a model's output and in identifying potential biases or reliance on spurious features \cite{14}.

Attention mechanisms, which are integral to transformer architectures but can also be incorporated into other neural network designs, provide an alternative pathway to explainability. By examining attention weights, it is possible to infer which segments of the input ECG signal the model prioritizes when forming a prediction \cite{15}. More sophisticated approaches have explored hierarchical attention mechanisms that aim to distinguish between features at different scales, such as beat-level morphology versus rhythm-level patterns, thereby offering explanations that are more aligned with clinical reasoning \cite{13, 16}. Nevertheless, many of these methods provide post-hoc explanations or rely on generic attention patterns, rather than integrating domain-specific cardiac knowledge directly into the attention computation process, which could lead to more inherently interpretable and clinically relevant models.

\subsection{Pattern Recognition in Cardiac Signals}

The concept of recognizing discrete, meaningful patterns within ECG signals is fundamental to clinical cardiology. For decades, clinicians have relied on identifying specific waveforms (e.g., P wave, QRS complex, T wave), intervals (e.g., PR, QT intervals), and morphological characteristics to diagnose cardiac conditions \cite{18}. Early computational approaches to ECG analysis attempted to codify this expert knowledge through rule-based systems and explicit feature engineering. However, these traditional methods often struggled with the inherent variability and noise present in real-world ECG data.

More recently, studies have shown that deep learning models, even when trained end-to-end, can implicitly learn to identify patterns that correspond to known ECG abnormalities, though these learned features often remain latent within the network's complex architecture \cite{4}. The idea of learned feature dictionaries or "electrophysiological motifs" has been explored, demonstrating that models can discover representations that align with clinically defined patterns, even without explicit guidance towards a predefined pattern vocabulary \cite{18}. Some research has proposed pattern-based approaches using learned templates for tasks like arrhythmia classification \cite{19}. While transformer-based feature extraction has also been investigated for ECG signals \cite{10}, a comprehensive framework that explicitly tokenizes ECGs into a vocabulary of physiological patterns and integrates this with domain-guided attention mechanisms remains an area ripe for exploration. Our work seeks to bridge this gap by explicitly framing ECG interpretation through a pattern vocabulary lens, aiming to enhance both performance and the clinical relevance of explanations.

\section{Methods}

\subsection{CardioPatternFormer Architecture Overview}
The CardioPatternFormer is a novel transformer-based architecture specifically designed for interpretable multi-label ECG classification. Components are carefully engineered to incorporate cardiological domain knowledge while leveraging the representational power of transformers. Fig. 1 presents an overview of the revised architecture. 

The CardioPatternFormer processes 12-lead ECG signals through several key stages:
\begin{enumerate}
    \item \textbf{Input Processing:} A \texttt{CardiacPatternTokenizer} transforms continuous signals into a sequence of embedded representations based on learned, multi-scale cardiac patterns. This is augmented by a \texttt{Multi-Resolution Temporal Encoding} layer that captures temporal relationships across various time scales.
    \item \textbf{Encoding:} A stack of Transformer encoder layers, utilizing specialized \texttt{Physiologically-Guided Attention} mechanisms (detailed in Sec~\ref{sec:physio_attention}), processes the embedded sequence to capture complex dependencies and contextual information.
    \item \textbf{Classification Heads:} The output of the encoder is fed into two parallel, complementary classification heads:
    \begin{itemize}
        \item An \texttt{ExplainableDiagnosticHead} generates class-specific attention maps alongside diagnostic predictions, providing interpretability.
        \item An \texttt{AdaptiveDiagnosticPooling} head uses a learnable relevance mechanism and uncertainty estimation for robust classification.
    \end{itemize}
    \item \textbf{Output Generation:} The logits from the two classification heads are fused using learnable weights (\texttt{diagnostic\_fusion}) to produce the final multi-label classification predictions. Corresponding attention maps and relevance scores from the heads provide insights into the classification decisions.
\end{enumerate}

\begin{figure}[htbp]
    \centering
    \includegraphics[width=0.7\linewidth]{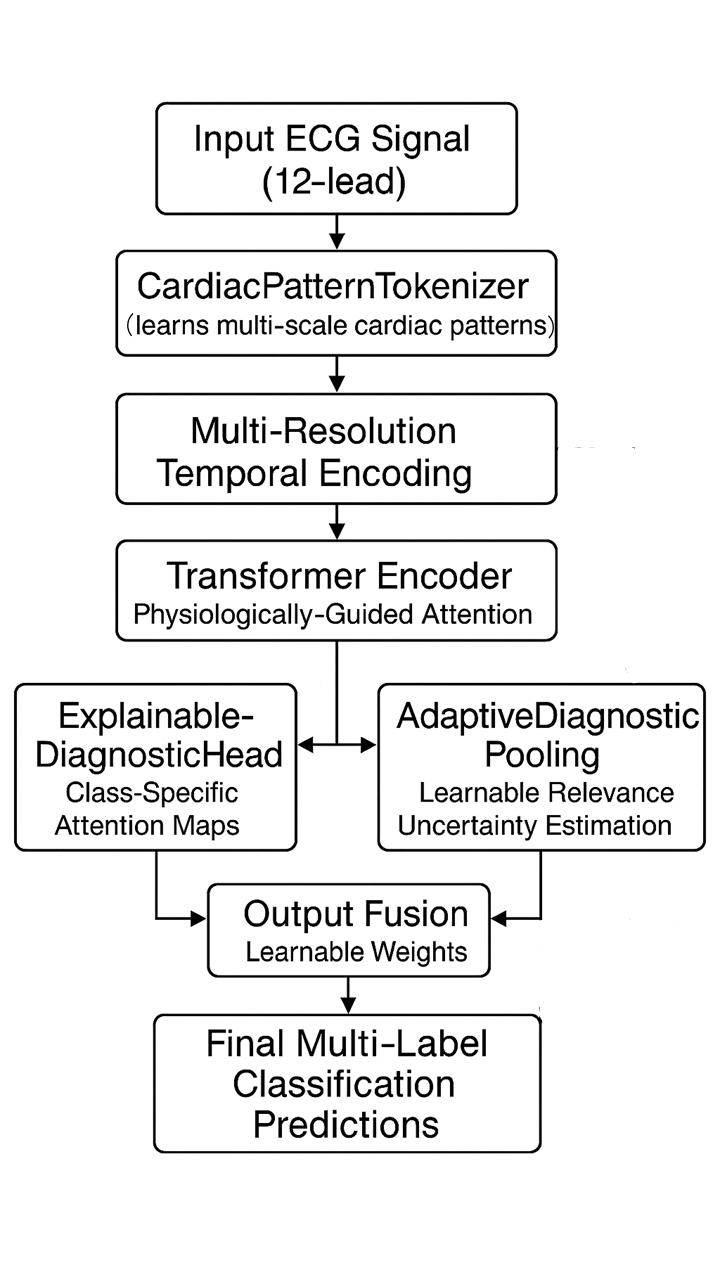} 
    \caption{Overview of the revised CardioPatternFormer architecture, illustrating the flow from input processing through encoding to the dual classification heads and output fusion.}
    \label{fig:arch_overview}
\end{figure}

\subsection{Cardiac Pattern Tokenizer}
Traditional approaches to ECG processing typically treat signals as continuous waveforms, applying generic convolutional operations without explicit consideration of underlying cardiac patterns. In contrast, our \texttt{CardiacPatternTokenizer} is inspired by the way cardiologists recognize distinct, meaningful patterns in ECG signals (e.g., P-waves, QRS complexes, ST-segments). The tokenizer employs multi-scale 1D convolutional layers (\texttt{nn.Conv1d}) with varying kernel sizes (e.g., 5, 9, 15, 25, 35) to detect patterns at different temporal resolutions, capturing both fine-grained morphological features and broader waveform characteristics. This multi-scale approach is critical for ECG interpretation, as diagnostic information exists at various temporal scales from narrow deflections to wider complexes and rhythm patterns spanning multiple heartbeats.

\begin{figure}[htbp]
    \centering
    \includegraphics[width=0.8\linewidth]{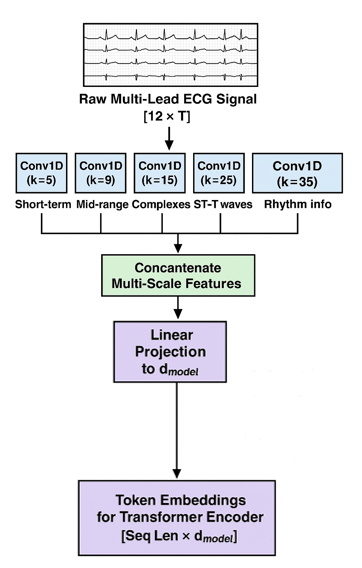} 
    \caption{Illustration of the multi-scale convolutional processing within the Cardiac Pattern Tokenizer.}
    \label{fig:tokenizer}
\end{figure}

The features extracted from each convolutional pathway are concatenated and projected to the model's embedding dimension ($d_{\text{model}}$). Unlike initializing with predefined waveforms, the convolutional kernels are initialized using standard methods (e.g., Kaiming normal) and learn relevant patterns directly from the data. A standard learned positional embedding is added to the resulting sequence of token embeddings to retain temporal order information. The tokenizer effectively transforms the raw multi-lead ECG into a sequence of rich feature vectors ready for the transformer encoder.

\subsection{Physiologically-Guided Attention}
\label{sec:physio_attention} 
Standard transformer self-attention mechanisms treat all input tokens uniformly, potentially overlooking known physiological relationships within ECG signals. To address this, our \texttt{PhysiologicallyGuidedAttention} mechanism integrates cardiac domain knowledge directly into the attention score calculation, enhancing the model’s focus on clinically relevant patterns.

\begin{figure}[htbp]
    \centering
    \includegraphics[width=0.8\linewidth]{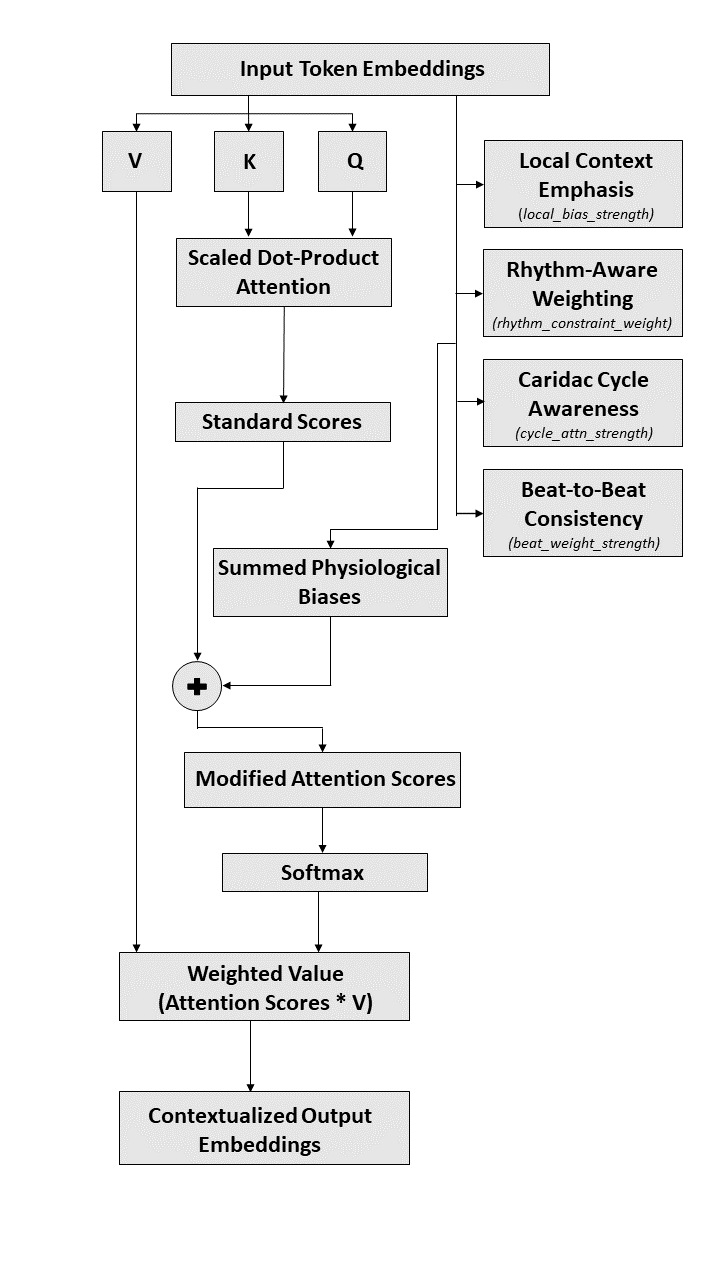} 
    \caption{Conceptual illustration of Physiologically-Guided Attention incorporating adaptable constraints, or an example of resulting attention patterns.}
    \label{fig:attention_viz}
\end{figure}

Instead of imposing fixed biases, this mechanism incorporates several physiological priors whose influence is controlled by learnable parameters, allowing the model to adapt the importance of each constraint during training. These adaptable constraints include:
\begin{enumerate}
    \item \textbf{Local Context Emphasis:} A relative positional bias is scaled by a learnable parameter (\texttt{local\_bias\_strength}). This encourages attention between temporally nearby tokens, reflecting the clinical principle that the interpretation of certain waveform segments (e.g., the ST segment or T-wave) often depends critically on the characteristics of the immediately preceding waveforms (e.g., the QRS complex), making local temporal context vital for morphological assessment.
    
    \item \textbf{Rhythm-Aware Weighting:} A periodic bias (e.g., cosine function based on relative position) is scaled by learnable, per-head weights (\texttt{rhythm\_constraint\_weight}). This allows different attention heads to potentially focus on different rhythmic patterns or periodicities relevant to arrhythmias or normal rhythms. Incorporating rhythm awareness is crucial as many cardiac conditions manifest primarily as deviations from normal sinus rhythm, requiring the model to effectively assess regularity and rate.
    
    \item \textbf{Cardiac Cycle Awareness:} A small convolutional network (\texttt{cycle\_detector}) processes the input embeddings to predict weights indicating the likelihood of specific cardiac cycle phases (e.g., QRS complex, T wave). These weights modulate the attention scores, scaled by another learnable parameter (\texttt{cycle\_attn\_strength}), guiding attention towards relevant phases. This allows the model to differentially weight information based on the specific phase of the P-QRS-T cycle, mimicking how clinicians focus on different waveform components to identify distinct abnormalities (e.g., P-waves for atrial issues, QRS for ventricular issues).
    
    \item \textbf{Beat-to-Beat Consistency:} A convolutional layer (\texttt{conv\_constraint}) operates on the attention patterns across the sequence length dimension, encouraging smoother or more consistent attention patterns between adjacent beats, scaled by a learnable \texttt{beat\_weight\_strength}. This encourages the model to recognize that for many stable conditions, waveform morphology is expected to be relatively consistent across consecutive beats, while abrupt changes in attention might signify transient events or noise.
\end{enumerate}
These learnable physiological constraints are added to the standard scaled dot-product attention scores before the softmax operation. This allows the model to leverage prior knowledge about ECG structure and dynamics while retaining the flexibility to adjust the influence of these priors based on the data and the specific diagnostic task. Fig.~\ref{fig:attention_viz} may illustrate the effect of these combined, adaptable constraints. 

\subsection{Multi-Resolution Temporal Encoding}
ECG interpretation requires understanding temporal relationships at multiple scales, from millisecond-level wave morphologies to second-level rhythm patterns. Standard positional embeddings capture order but not necessarily scale. Our \texttt{MultiResolutionTemporalEncoding} addresses this by learning distinct temporal embeddings at multiple resolutions simultaneously (e.g., using scales of 1, 2, 5, 10, 20 relative to the input sequence length).

For each resolution scale, a separate learnable embedding parameter (\texttt{nn.Parameter}) is defined. During the forward pass, the appropriate length of each resolution's embedding is selected and then interpolated (using \texttt{F.interpolate} with linear mode) to match the actual input sequence length. These interpolated embeddings, representing temporal information at different scales, are concatenated along the feature dimension. A final linear projection layer (\texttt{self.projection}) maps the concatenated multi-resolution temporal features back to the model's embedding dimension. This projected temporal encoding is then added to the input token embeddings (after the \texttt{CardiacPatternTokenizer}), providing the transformer layers with rich, multi-scale temporal context. This allows the model to adaptively leverage fine-grained temporal information for morphological analysis and broader temporal information for rhythm analysis.

\subsection{Explainable Diagnostic Classification}
Following the transformer encoder stack, CardioPatternFormer
employs two distinct classification heads in parallel, whose
outputs are ultimately fused:
\begin{enumerate}
    \item \textbf{\texttt{ExplainableDiagnosticHead}:} This head aims to
    provide interpretability alongside classification. For each
    of the \texttt{num\_classes} diagnostic categories, it uses a sep-
    arate small attention network (\texttt{class\_attentions})
    to compute attention scores over the encoder’s output sequence. These scores generate class-specific attention maps, indicating which parts of the ECG
    signal were most influential for predicting that spe-
    cific class. The weighted context vector for each
    class is then passed through a dedicated predictor
    (\texttt{class\_predictors}) to generate the class logit.
    This head returns both the logits and the corresponding
    attention maps (\texttt{explanation\_maps}).
    
    \item \textbf{\texttt{AdaptiveDiagnosticPooling}:} This head
    offers an alternative classification approach focused
    on relevance and uncertainty. It first uses a
    \texttt{relevance\_detector} (a linear layer followed
    by a sigmoid) to calculate a relevance score for each
    time step in the encoder output sequence. These scores
    weight the importance of each time step’s features.
    A weighted average pooling is performed, using the
    relevance scores as weights, to obtain a single context
    vector representing the most relevant parts of the signal.
    This pooled vector is fed through a \texttt{classifier}
    to produce diagnostic logits. Additionally, the pooled
    vector is passed to an \texttt{uncertainty\_estimator}
    head to predict an uncertainty score (between 0 and
    1) for each class prediction. This head returns logits,
    relevance maps (\texttt{relevance}), and uncertainty scores
    (\texttt{uncertainty}).
\end{enumerate}
The final classification logits are obtained by combin-
ing the logits from the \texttt{ExplainableDiagnosticHead}
and \texttt{AdaptiveDiagnosticPooling} head using learned
scalar weights (\texttt{diagnostic\_fusion}).

\begin{figure*}[htbp] 
    \centering
    \includegraphics[width=0.7\linewidth]{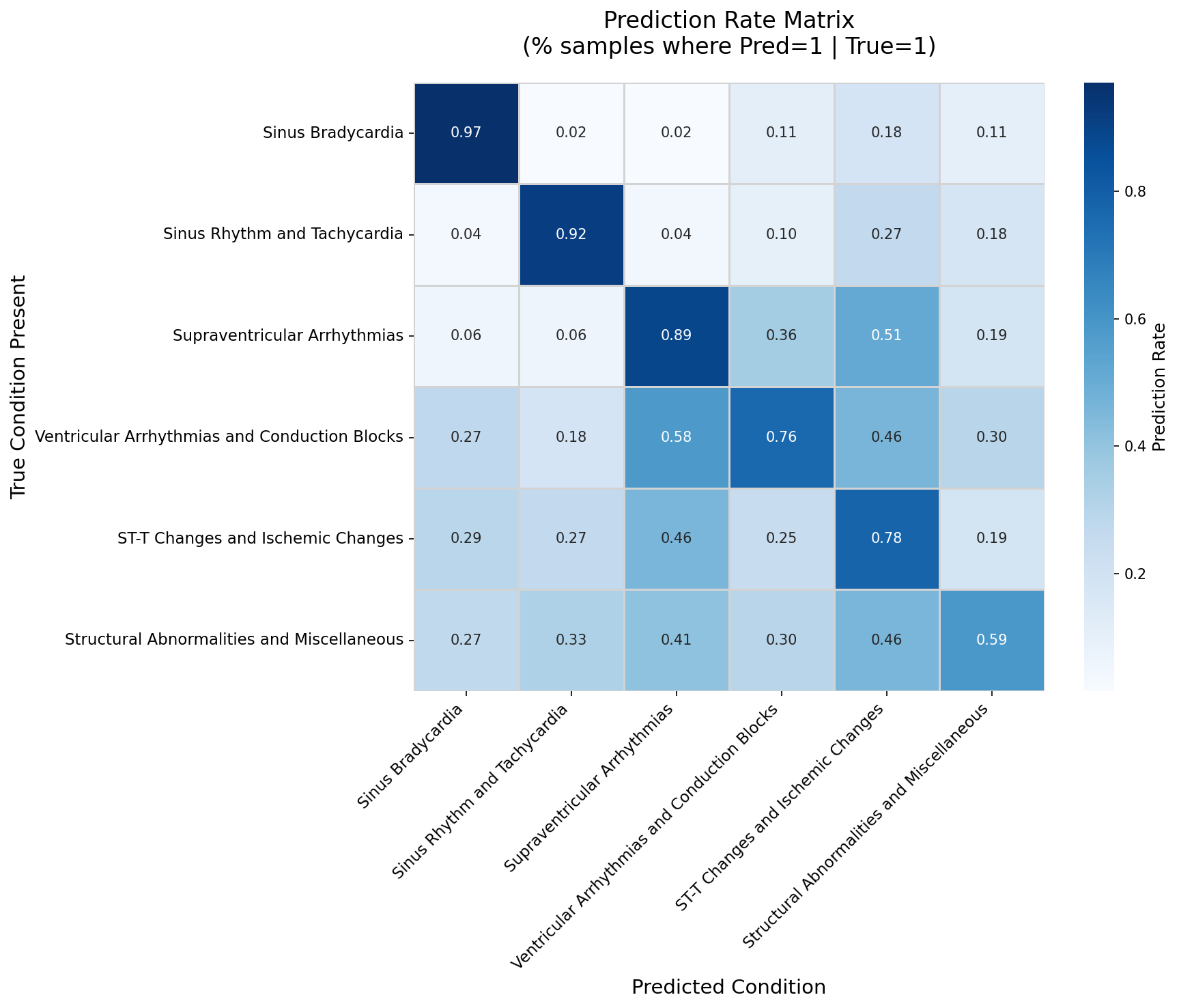} 
    \caption{Heatmap illustrating the Prediction Rate Matrix on the test set. Rows indicate the true condition present, columns indicate the predicted condition, and cell values represent the rate P(Pred=1 | True=1). The diagonal elements highlight the recall (sensitivity) achieved for each diagnostic category.}
    \label{fig:comp_viz}
\end{figure*}

\subsection{Loss Function}
To train the model effectively for the multi-label classification task, we utilize a specialized loss function strategy that incorporates domain knowledge and addresses common challenges in medical datasets. The primary component is an \textbf{Enhanced Physiological Focal Loss} function. This builds upon the standard Focal Loss concept by focusing the training process on more challenging diagnostic examples that the model initially classifies poorly. Furthermore, it mitigates the issue of class imbalance, which is common in medical datasets, by applying weights that give more importance to underrepresented diagnostic categories. This classification loss can also optionally incorporate prior knowledge about the physiological relationships between different cardiac conditions, encouraging the model to make predictions consistent with known clinical co-occurrences.

In addition to the primary classification loss, we incorporate an optional \textbf{Attention Diversity Loss}. When used (controlled by a hyperparameter weight), this auxiliary loss encourages the explanation mechanisms within the model (specifically, the class-specific attention maps) to focus on distinct regions of the ECG signal for different diagnoses, where appropriate. This promotes more varied and potentially more insightful explanations for each predicted condition.

The overall training objective combines the primary classification loss and the optional attention diversity loss, using hyperparameters to balance their respective contributions. This carefully designed loss strategy guides the model towards learning accurate, robust, and interpretable ECG classifications.

\section{Experiments and Results}

\subsection{Dataset and Preprocessing}
We evaluated CardioPatternFormer on the publicly available Chapman-Shaoxing ECG dataset, which contains 12-lead ECG recordings from 10,247 patients. Each recording is accompanied by cardiologist-annotated diagnostic labels conforming to the SCP-ECG standard. For model training and evaluation, we followed common practice by mapping the numerous specific diagnostic codes into six clinically relevant, broader categories:
\begin{enumerate}
    \item Sinus Bradycardia
    \item Sinus Rhythm and Tachycardia
    \item Supraventricular Arrhythmias
    \item Ventricular Arrhythmias and Conduction Blocks
    \item ST-T Changes and Ischemic Changes
    \item Structural Abnormalities and Miscellaneous
\end{enumerate}
This grouping results in a multi-label classification task where each ECG can belong to one or more categories. The dataset exhibits a natural imbalance across these categories, as detailed in Table~\ref{tab:dataset_dist}, motivating the use of specialized loss functions described in Section~III-F. 

\begin{table}[htbp]
\caption{Dataset Characteristics and Diagnostic Category Distribution}
\label{tab:dataset_dist}
\begin{center}
 \begin{tabular}{p{5cm}cccc} 
 \hline
 \textbf{Diagnostic Category} & \textbf{Prevalence (\% or Count)} \\
 \hline
 Sinus Bradycardia & 33.07\% \\
 Sinus Rhythm and Tachycardia & 33.12\% \\
 Supraventricular Arrhythmias & 31.11\% \\
 Ventricular Arrhythmias and Conduction Blocks & 14.24\% \\
 ST-T Changes and Ischemic Changes & 29.29\% \\
 Structural Abnormalities and Miscellaneous & 11.70\% \\
 \hline
 \end{tabular}
\end{center}
\end{table}

The original ECG signals were recorded at 500 Hz, typically spanning 10 seconds ($\approx$5,000 time points). Our preprocessing pipeline involved several steps applied to each recording:
\begin{itemize}
    \item \textbf{Filtering:} A fourth-order Butterworth bandpass filter between 0.5 Hz and 45 Hz was applied to remove baseline wander and high-frequency noise.
    \item \textbf{Resampling:} Signals were downsampled from 500 Hz to a target frequency of 100 Hz, reducing the sequence length to approximately 1,000 time points while preserving clinically relevant frequencies.
    \item \textbf{Normalization:} Each of the 12 leads was independently normalized to have zero mean and unit variance.
\end{itemize}
This preprocessing resulted in 12-lead ECG tensors of shape (12, $\approx$1000) along with corresponding multi-label diagnostic vectors and calculated parameter vectors for input into the CardioPatternFormer model.

\subsection{Experimental Setup}
We implemented the CardioPatternFormer model using PyTorch. The model architecture utilized an embedding dimension ($d_{\text{model}}$) of 256, 8 attention heads within the \texttt{PhysiologicallyGuidedAttention} layers, and a stack of 4 transformer encoder layers. The \texttt{CardiacPatternTokenizer} used a pattern vocabulary size of 16 (\texttt{num\_patterns}).

The model was trained using the AdamW optimizer with an initial learning rate of $5 \times 10^{-5}$ and a weight decay of 0.01. A \texttt{ReduceLROnPlateau} learning rate scheduler was employed, monitoring the validation macro F1-score, reducing the learning rate by a factor of 0.2 if no improvement was observed for 3 epochs. The primary classification loss function was the Enhanced Physiological Focal Loss (detailed in Section~III-F), configured with parameters $\alpha=0.5$ and $\gamma=2.0$. 

We utilized a 5-fold cross-validation strategy on approximately 85\% of the full dataset designated for training and validation. The remaining $\approx$15\% was held out as a final test set. Within each fold of the cross-validation, the data was split into training and validation sets. Training was performed for a maximum of 30 epochs, using a batch size of 16 and gradient accumulation over 2 steps, resulting in an effective batch size of 32. Automatic mixed-precision (AMP) training was enabled to optimize computational efficiency and memory usage. We implemented early stopping with a patience of 5 epochs based on the validation macro F1-score to prevent overfitting.

All reported performance metrics in Section~\ref{sec:perf_results} onwards are based on the evaluation of the best model (selected based on the highest validation macro F1-score during cross-validation) on the held-out test set. 

\subsection{Classification Performance}
\label{sec:perf_results}
CardioPatternFormer demonstrated strong multi-label classification performance on the held-out Chapman-Shaoxing test set. The model achieved an overall Hamming accuracy of 0.9184 and a macro F1-score of 0.8019. Table~\ref{tab:overall_perf} summarizes the key macro-averaged performance metrics on the test set.

\begin{table}[htbp]
\caption{Overall Performance on Chapman-Shaoxing Test Set}
\label{tab:overall_perf}
\begin{center}
\begin{tabular}{lc}
\hline
\textbf{Metric} & \textbf{Value} \\
\hline
Accuracy (Hamming) & 0.9184 \\
Macro Precision & 0.7896 \\
Macro Recall    & 0.8184 \\
Macro F1-Score  & 0.8019 \\
Macro AUC       & 0.9437 \\
\hline
\end{tabular}
\end{center}
\end{table}

The high macro AUC value (0.9437) indicates robust discriminative ability across all diagnostic categories, suggesting the model effectively distinguishes between positive and negative cases even with the dataset's inherent class imbalance.

Table~\ref{tab:per_class_perf} presents the detailed per-class performance metrics, evaluated using the optimal thresholds determined during cross-validation. These results reveal variations in performance across the different diagnostic groups.

\begin{table}[htbp] 
\caption{Per-Class Performance Metrics on Test Set}
\label{tab:per_class_perf}
\begin{center}
\begin{tabular}{p{1.7cm}ccccc}
\hline
\textbf{Condition} & \textbf{Precision} & \textbf{Recall} & \textbf{F1 Score} & \textbf{AUC} & \textbf{Threshold} \\
\hline
Sinus Bradycardia                           & 0.9475 & 0.9705 & 0.9588 & 0.9950 & 0.396 \\
Sinus Rhythm and Tachycardia                & 0.9639 & 0.9179 & 0.9404 & 0.9869 & 0.554 \\
Supraventricular Arrhythmias                & 0.9408 & 0.8919 & 0.9157 & 0.9719 & 0.446 \\
Ventricular Arrhythmias and Conduction Blocks & 0.6691 & 0.7647 & 0.7137 & 0.9414 & 0.386 \\
ST-T Changes and Ischemic Changes           & 0.7585 & 0.7785 & 0.7684 & 0.9190 & 0.416 \\
Structural Abnormalities and Miscellaneous  & 0.4576 & 0.5870 & 0.5143 & 0.8481 & 0.406 \\
\hline
\end{tabular}
\end{center}
\end{table}

Consistent with clinical intuition, the model performs exceptionally well on distinct rhythm categories like Sinus Bradycardia (F1=0.9588) and differentiating normal Sinus Rhythm/Tachycardia (F1=0.9404). Performance remains strong for Supraventricular Arrhythmias (F1=0.9157). As expected, performance is moderately lower for the more heterogeneous and potentially subtler categories of Ventricular Arrhythmias/Conduction Blocks (F1=0.7137) and ST-T Changes/Ischemic Changes (F1=0.7684). The lowest F1-score is observed for Structural Abnormalities/Miscellaneous (F1=0.5143), likely reflecting the combined challenge of diagnostic subtlety from ECG alone and lower prevalence in the dataset. Despite these variations, the per-class AUC values remain high ($\ge$0.84 for all classes ), indicating good underlying discriminative capability across all conditions. This performance gradient generally aligns with clinical experience regarding the relative difficulty of diagnosing these conditions solely from ECG signals.
\par 
To contextualize CardioPatternFormer's performance, Table~\ref{tab:simplified_comparison} compares its results with selected studies utilizing the Chapman-Shaoxing dataset. However, readers should note that direct comparison is inherently challenging due to significant variations in task definitions (e.g., specific arrhythmias vs. broad categories), reported metrics, and evaluation protocols across different publications. Therefore, the table serves primarily as a contextual reference rather than a direct leaderboard.

\begin{table*}[htbp] 
\caption{Simplified Performance Comparison with Selected Related Works (Chapman-Shaoxing Dataset Context).}
\label{tab:simplified_comparison} 
\begin{center}
\begin{tabular}{p{4cm} p{4.5cm} p{4.5cm} c} 
\toprule 
\textbf{Model} & \textbf{Architecture (Brief)} & \textbf{Task Focus (Leads, Classes / Type)} & \textbf{Accuracy Results (\% or Score)} \\ 
\midrule

\textbf{CardioPatternFormer } & \textbf{Transformer (Pattern Tok. + Guided Attn)} & \textbf{12L / 6 Broad Multi-label Cats.} & \textbf{91.84 (Hamming Acc.)} \\ 
\midrule

Bimodal CNN  & Dual Inception CNN (Img + Scalogram) & 12L / 11 Rhythms & $\approx$95.7 (Accuracy) \\
CNN-BiLSTM-BiGRU  & CNN+BiLSTM+BiGRU+Attn & Lead II / 7 Rhythms & $\approx$98.6 (Accuracy) \\
Evo CNN Trees  & Evo. CNN-Tree Fusion & 12L / 11 Rhythms & $\approx$97.6 (Avg. Accuracy) \\
CNN Teacher+Student  & CNN (Knowledge Distill.) & 12L / 11 Rhythms & 97.55 (Accuracy - Student) \\
CMA Classifier  & Custom Feature-Image Classif. & 12L / 5 Rhythms & 99.76 (Accuracy) \\
GCN-WMI  & Graph ConvNet (15-layer) & 12L / Mult. Rhythms & 99.82 (Accuracy) \\

\bottomrule
\end{tabular}
\end{center}
\end{table*}

\subsection{Interpretability Analysis}
A key advantage of the CardioPatternFormer architecture is its capacity for interpretability, primarily facilitated by the \texttt{ExplainableDiagnosticHead}. This component generates class-specific attention maps (\texttt{explanation\_maps}) alongside its predictions. These maps are designed to highlight the temporal regions within the input ECG signal that were most influential in reaching a specific diagnostic conclusion.

\begin{figure}[htbp]
    \centering
    \includegraphics[width=\linewidth]{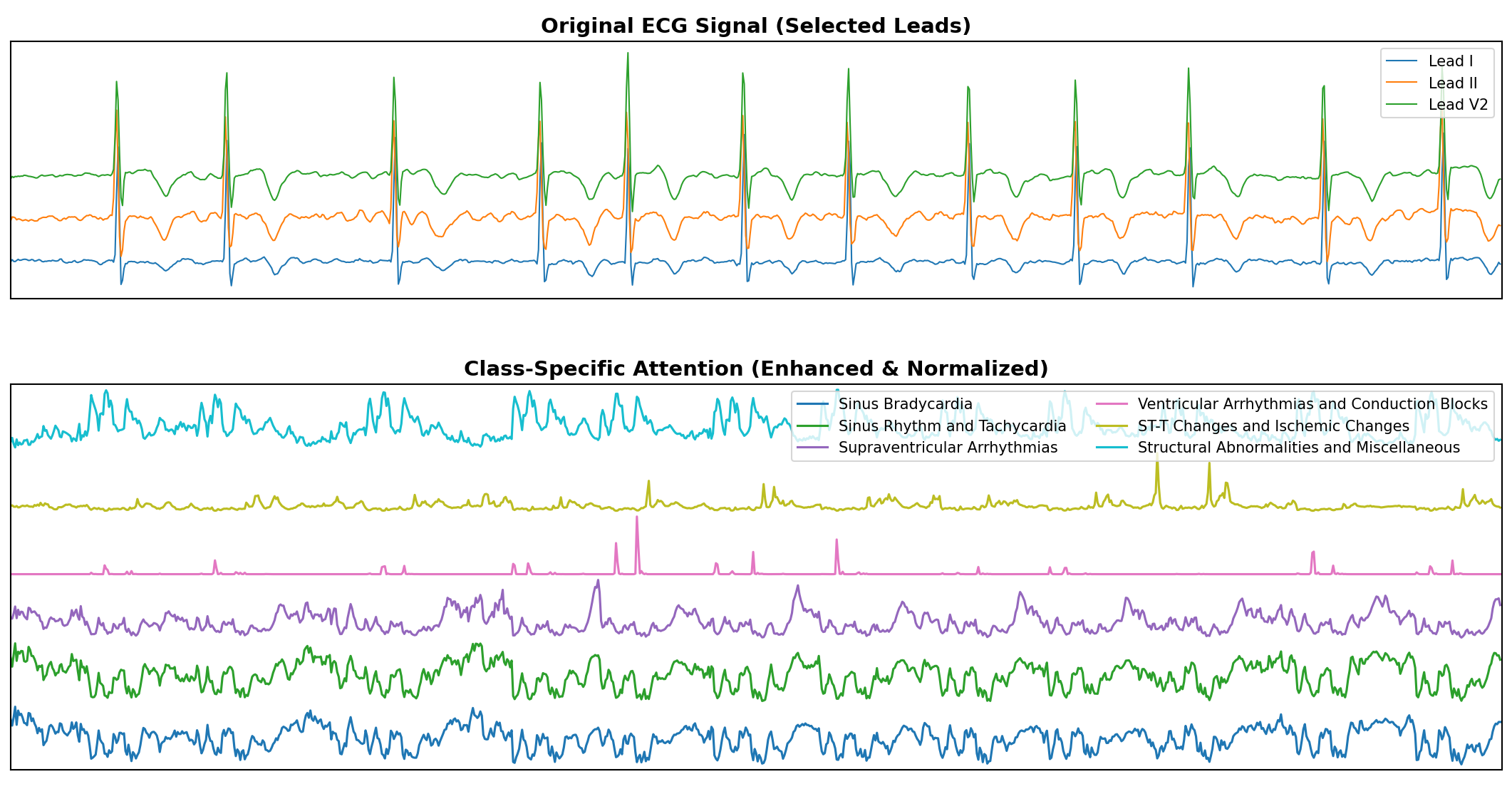} 
    \caption{Example class-specific attention maps from the \texttt{ExplainableDiagnosticHead} for a representative test sample. Top panel: Selected input ECG leads (e.g., I, II, V2). Bottom panel: Corresponding normalized attention weights over time for different diagnostic classes, illustrating the model's temporal focus for each potential diagnosis.}
    \label{fig:attention_example} 
\end{figure}

Fig.~\ref{fig:attention_example} provides an example of these class-specific attention maps. 

By visualizing these attention scores, typically overlaid on the ECG signal (as shown in Fig.~\ref{fig:attention_example}), it is possible to gain insights into the model's decision-making process. The goal is for these highlighted regions to correspond to physiologically relevant waveforms or patterns associated with the predicted condition. This mechanism aims to enhance the model's clinical utility by providing explanations that can be evaluated against established electrophysiological knowledge, thereby fostering trust and potentially aiding clinical review.

Furthermore, comparing attention patterns across different predicted diagnoses can illustrate how the model potentially uses distinct ECG features to differentiate between various conditions, which is essential for accurate multi-label classification.

\subsection{Analysis of Complex Cases}
\begin{figure}[!htb]
    \centering
    \includegraphics[width=\linewidth]{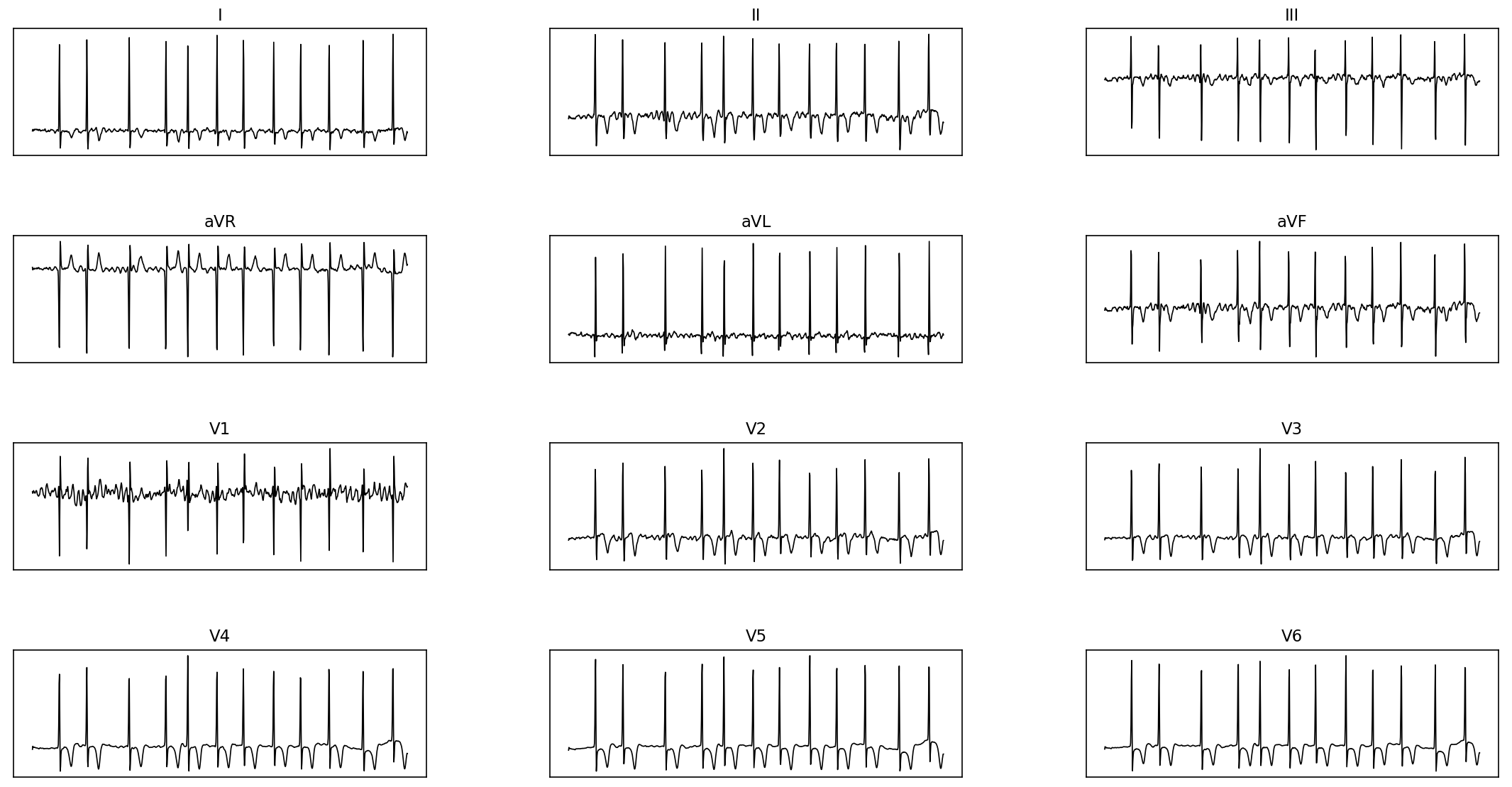}
    \caption{A representative 12-lead ECG signal from a patient presenting with multiple simultaneous cardiac conditions. The signal spans 10 seconds and is sampled at 100 Hz. 
    }
    \label{fig:complex_case_ekg}
\end{figure}

\begin{figure*}[!htb]
    \centering
    \includegraphics[width=\linewidth]{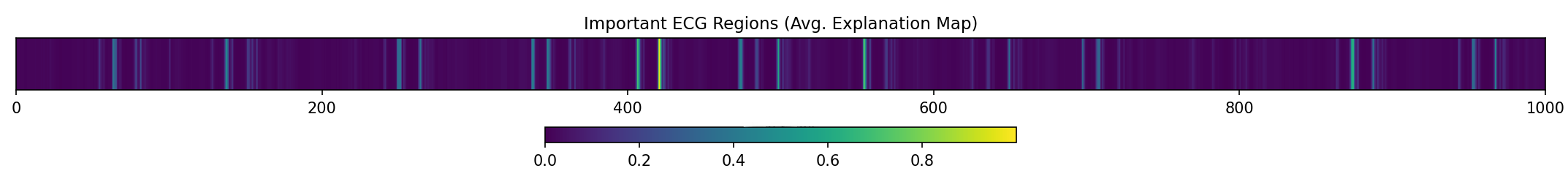} 
    \caption{Average attention map of CardioPatternFormer for the ECG signal in Fig.~\ref{fig:complex_case_ekg}, highlighting the regions of increased model focus. The color intensity represents the average attention weight given by the model across all detected patterns, with brighter colors (yellow/green) indicating higher attention.}
    \label{fig:complex_case_attention}
\end{figure*}

Real-world ECG interpretation often involves patients with multiple simultaneous cardiac conditions. To assess CardioPatternFormer's performance in such challenging scenarios, we examined its predictions and explanations on test cases exhibiting multiple ground-truth diagnoses. Fig.~\ref{fig:complex_case_ekg} presents a representative 12-lead ECG example from such a complex case. The detailed multi-label classification results for this specific case are provided in Table~\ref{tab:complex_case_predictions}, while the corresponding average attention map, highlighting the most influential signal regions for the model's overall prediction, is shown in Fig.~\ref{fig:complex_case_attention}.

\begin{table}[!htb]
    \caption{Multi-label Classification Results for a Complex ECG Case with Multiple Co-occurring Conditions.}
    \label{tab:complex_case_predictions}
    \centering
    \begin{tabular}{p{3cm} c c r} 
        \toprule
        \textbf{Cardiac Condition} & \textbf{Ground Truth} & \textbf{Prediction} & \textbf{Probability} \\ 
        \midrule
        Sinus Bradycardia & & & 0.028 \\
        Sinus Rhythm and Tachycardia & & & 0.066 \\
        Supraventricular Arrhythmias & \checkmark & \checkmark & 0.924 \\
        Ventricular Arrhythmias and Conduction Blocks & & & 0.407 \\
        ST-T Changes and Ischemic Changes & \checkmark & \checkmark & 0.729 \\
        Structural Abnormalities and Miscellaneous & & & 0.295 \\
        \bottomrule
    \end{tabular}
\end{table}

In representative examples, the model demonstrated the ability to identify several co-occurring conditions correctly. Examination of the corresponding class-specific attention or relevance maps suggests the model often leverages distinct signal features or time windows when predicting different concurrent diagnoses, aligning with the expectation that different conditions may have unique ECG manifestations even when present simultaneously. This capability is crucial for providing comprehensive and clinically useful interpretations in complex cases.

\subsection{Lead Ablation Study}
To evaluate the model's robustness and the relative importance of different ECG leads for classification performance, we conducted a lead ablation study. Using the best-performing model checkpoint obtained from cross-validation, we systematically evaluated its performance on the test set while providing only specific subsets of the original 12 leads as input. Unavailable leads were masked by zero-padding their corresponding input channels before entering the \texttt{CardiacPatternTokenizer}. We tested various configurations, including all single leads, standard clinical subsets (e.g., limb leads only, precordial leads only), and other combinations. Performance was measured using macro F1-score and AUC for classification, applying the optimal thresholds determined previously.

The results demonstrate a clear dependency on the number of available leads for optimal classification performance. Performance degrades significantly when using fewer leads compared to the full 12-lead configuration. Table~\ref{tab:ablation_results} summarizes the macro F1-scores achieved for representative lead subsets.

\begin{table}[htbp]
\caption{CardioPatternFormer Classification Performance (Macro F1-Score) with Reduced Lead Sets}
\label{tab:ablation_results}
\begin{center}
\begin{tabular}{p{2.7cm}ccccc}
\hline
\textbf{Lead Configuration} & \textbf{Leads Included} & \textbf{Macro F1-Score} \\
\hline
Single Lead (Best)        & II                      & 0.3602 \\
Single Lead (Worst)       & V3                      & 0.2828 \\
\textit{Average Single Lead (approx)} & \textit{(Range: $\approx$0.28 - 0.36)} & \textit{($\approx$0.34)} \\
Limb Leads Only           & I, II, III, aVR, aVL, aVF & 0.6122 \\
Precordial Leads Only     & V1, V2, V3, V4, V5, V6  & 0.5809 \\
Example Reduced Set (4 Leads) & I, II, V1, V5       & 0.5639 \\
Full 12-Lead Set          & All                     & \textbf{0.8019} \\
\hline
\end{tabular}
\end{center}
\end{table}

As shown in Table~\ref{tab:ablation_results}, while single leads provide limited diagnostic information (macro F1 scores ranging from approximately 0.28 to 0.36), combining leads significantly improves performance. Standard 6-lead configurations (limb or precordial) achieve intermediate results, while the full 12-lead set provides substantially better overall classification accuracy. Among the single leads, Lead II achieved the highest individual macro F1-score in our test, aligning with its common use in rhythm assessment, although performance with any single lead was considerably lower than multi-lead configurations. Analysis also revealed variations in the importance of specific leads for classifying different diagnostic categories.

From this analysis, we observe the model's adaptability to varying lead inputs, confirming the expected performance trade-offs, and gain quantitative metrics regarding the importance of individual leads for the multi-label classification task.

\section{Discussion}
CardioPatternFormer represents a significant step towards automated ECG interpretation systems that are not only accurate but also interpretable and aligned with clinical reasoning. By framing ECG analysis through the lens of pattern recognition and developing specialized architectural components like the \texttt{CardiacPatternTokenizer}, \texttt{MultiResolutionTemporalEncoding}, and particularly the \texttt{PhysiologicallyGuidedAttention} with learnable constraints, our approach aims to bridge the gap between black-box deep learning models and the nuanced process of clinical diagnosis. The model demonstrated strong classification performance on the challenging multi-label Chapman-Shaoxing dataset, achieving a macro F1-score of 0.8019 and a macro AUC of 0.9437 on the held-out test set. Analysis of per-class performance revealed a clinically plausible gradient: the model excelled at identifying conditions with distinct ECG patterns like Sinus Bradycardia (F1=0.9588) and Sinus Rhythm/Tachycardia (F1=0.9404), while performance was lower, yet still substantial (AUC $\ge$ 0.84), for more heterogeneous or subtle categories like Structural Abnormalities/Miscellaneous (F1=0.5143). This suggests the model learns meaningful diagnostic representations.

A key contribution of this work is the enhanced interpretability offered through class-specific attention maps generated by the \texttt{ExplainableDiagnosticHead}. Our qualitative analysis indicated that the model often focuses on physiologically relevant ECG regions corresponding to the predicted diagnosis, such as specific waveform segments or intervals pertinent to different conditions. This capability, visualized through attention maps, provides valuable transparency into the model's decision process, addressing a major limitation of previous deep learning approaches and potentially fostering greater clinical trust and utility. Furthermore, the inclusion of an \texttt{AdaptiveDiagnosticPooling} head provides complementary diagnostic information, including uncertainty estimates that correlate with prediction accuracy and could help prioritize challenging cases for expert review in clinical workflows. The overall pattern-based approach, combining multi-scale feature extraction with attention mechanisms that adaptively incorporate domain knowledge, appears effective for handling complex multi-label ECG classification.

Despite promising results, several limitations should be acknowledged. The current architecture relies on fixed-length inputs ($\approx$1000 time points after downsampling), which might limit its ability to capture very long-term dependencies relevant to certain intermittent arrhythmias. While the Chapman-Shaoxing dataset is relatively large, generalizability to patient populations with different demographics or ECG acquisition protocols requires further investigation. Class imbalance remains a challenge; although mitigated by the specialized loss function, performance on underrepresented conditions is still lower, highlighting the need for larger, more diverse datasets. Furthermore, a significant limitation in this study relates to the auxiliary task of physiological parameter prediction; the automated method used for calculating ground-truth parameters (particularly PR, QRS, QTc intervals) during data preprocessing was insufficient, leading to unreliable parameter prediction results. Future work necessitates implementing robust ECG delineation algorithms for accurate ground-truth parameter extraction before this component can be meaningfully evaluated. Our evaluation focused on classification accuracy and interpretability visualizations; comprehensive clinical validation, including usability studies and comparisons with existing decision support tools, is essential before deployment. Finally, while attention maps indicate \textit{where} the model focuses, they don't fully elucidate the \textit{physiological reasoning}, representing an ongoing challenge in deep learning explainability.

There are several interesting directions that could be explored in future work. One idea is to improve the \texttt{CardiacPatternTokenizer} so it can learn a more structured or layered vocabulary of cardiac patterns, which might help make the model both more accurate and easier to interpret. Another area worth looking into is how to better capture long-range temporal relationships, especially since this could be important for analyzing more complex heart rhythms. It’s also important to address the current challenges in parameter prediction, possibly by developing better ways to calculate ground-truth values. Beyond technical improvements, running clinical evaluation studies would be a key step in understanding how CardioPatternFormer might actually perform in real-world healthcare settings. Lastly, combining ECG analysis with other types of clinical data like patient history, lab tests, or imaging could make the system more useful in supporting broader diagnostic decisions.
\section{Conclusion}
CardioPatternFormer offers a promising approach to interpretable, automated ECG classification by leveraging a pattern-recognition perspective and integrating domain knowledge through specialized transformer components. Our results demonstrate strong classification performance across a diverse set of cardiac conditions on the Chapman-Shaoxing dataset, with enhanced transparency provided by class-specific attention mechanisms that highlight physiologically relevant signal regions. The ability to generate explanations aligned with clinical reasoning, alongside robust classification, marks progress towards AI tools that can function synergistically with clinicians. While further development and validation are essential, particularly concerning robust physiological parameter extraction and prospective clinical studies, CardioPatternFormer establishes a valuable framework for advancing ECG analysis. For AI tools like CardioPatternFormer to be successfully integrated into clinical practice, prioritizing transparency and alignment with clinical workflows, as we have aimed to do, is crucial. Such interpretable systems have the potential to foster clinician trust and enable effective human-AI collaboration, ultimately leading to improved patient care.

\end{document}